\documentclass[twocolumn,english,prx]{revtex4-1}
\usepackage{color}
\usepackage[utf8]{inputenc}
\usepackage{graphicx}
\usepackage[T1]{fontenc}
\usepackage{float}
\usepackage{natbib}
\usepackage{textcomp}
\usepackage{amsmath}
\usepackage{amssymb}
\usepackage{graphicx}
\usepackage{esint}
\usepackage{natbib}
\usepackage{xcolor}
\usepackage{multirow}
\usepackage{ulem}

\begin{document}
\title{Annular confinement for electrons on liquid helium}

\author{Bart\l{}omiej Szafran}

\affiliation{AGH University of Science and Technology, Faculty of Physics and
Applied Computer Science,\\
 al. Mickiewicza 30, 30-059 Kraków, Poland}

\begin{abstract}
We discuss the annular electron confinement on the liquid helium surface induced by a submerged tubular electrode. For a shallow liquid layer 
 the resulting potential  has a minimum  off the symmetry axis of the electrode. 
The ground-state angular momentum transitions that are driven by the external magnetic field 
can be resolved when
 the confinement radii in the first and second Rydberg subbands of the vertical quantization
are different, e.g., for  submersion depth of the tube comparable to its radius.
Then, discontinuities in the main microwave absorption line appear with the period that corresponds
to the subsequent magnetic flux quanta passing across the area within the ground-state confinement radius. 
\end{abstract}
\maketitle
\section{Introduction}

The surface of liquid helium can serve as an ultraclean substrate for the two-dimensional electron gas \cite{g74}. Electrons at the vacuum side are bound by the image charges that they
induce in the weakly dielectric liquid \cite{g74,g76,co02}.
The vertical quantization of  electron motion
in the Coulomb potential of the image \cite{g76,co02}
 gives rise to
 Rydberg states of hydrogen-like spectrum \cite{g74,g76,co02,kawakami21}. 
The pristine nature of the substrate allowed for the first observation \cite{w792} of the Wigner
crystallization at low electron gas density  \cite{w791,haq03}.
For the study of electron states at the surface, microwave spectroscopic techniques \cite{g76} are  developed \cite{is,yunusowa,zadorozhko,kk,lambs,kawakami19} with
recent advances on intersubband absorption \cite{kk}, excitations of high-energy Rydberg states \cite{kawakami21}, relaxation times \cite{yunusowa}, and coupling with the superconducting 
 resonance cavity \cite{trap,yang,kool}.

In addition to the vertical confinement 
a lateral one can be introduced by gates defined at the sides of the container or under the surface of helium.
The submerged electrodes
locally strengthen the image charge potential \cite{1D,Pl,Dy,Dyprb,Ly,Da,PE,Gl,sab08}. The gates
are considered for e.g., one-dimensional channels \cite{1D} or 0D localized states \cite{Pl,Dy,Dyprb,Ly,Da,Gl} for  quantum information storage and processing \cite{Gl,sab08}. 
Recently, the coupling of Rydberg states of vertical quantization with the lateral confinement of the Landau levels by the in-plane magnetic field
has been studied \cite{yunusowa,zadorozhko}  as a model of an atom interacting with an oscillator potential \cite{zadorozhko}.

In this paper, we study look for the annular confinement of  Rydberg states by a submerged
tube-shaped electrode. In open
ring-shaped solid-state devices \cite{webb,lb,fuhrer} the Aharonov-Bohm \cite{ab} effect is manifested by conductance oscillations that exhibit periodicity with the flux of the magnetic field threading 
the ring. On the other hand, for closed circular quantum rings \cite{qr} the ground-state undergoes angular
momentum transitions as subsequent magnetic flux quanta fit inside the ring  \cite{fuhrer,fominprl}. 
We discuss the possibility of observation of these angular momentum transitions
in the microwave absorption spectra of electron states localized above the tube electrode.
Recently, the formation of quantum rings on the surface of a semiconductor by single-atom engineering has been reported \cite{pham}.

\section{Theory}

The confinement of the electron gas at the liquid helium surface is due to the dielectric constant
discontinuity at the liquid/vacuum interface \cite{g74,g76,co02}.
Experiments on microwave absorption by electrons localized at the liquid helium surface are performed
in a parallel capacitor configuration \cite{kawakami21,is,yunusowa,zadorozhko} introducing an electric field that neutralizes the surface electron charge and tunes the transition energies between Rydberg states \cite{yunusowa,kawakami21}. 
A cross-section of the model system is depicted in Fig. 1. The lower plate contains a circular tube
electrode protruding towards the helium surface 
with an inner radius $R_1$ and $R_2$.  The lower plate is submerged in liquid helium of depth $H$. 
The tube electrode reaches depth $d$ beneath the helium surface.
For the bulk of this work, we consider the model system with $R_1=400$nm, $R_2=500$nm and 
$H=1.2 \mu$m. Discussion of the geometrical parameters for absorption spectra is provided below.

\begin{figure}[htbp]
\centering
\includegraphics[height=0.28\textwidth]{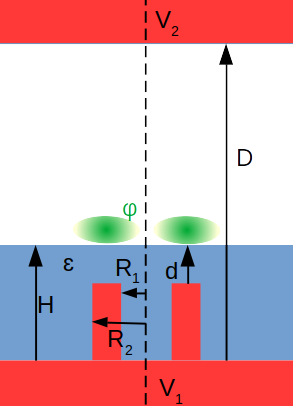} \put(-80,110) {$z\uparrow$}\put(-80,100) {$x\rightarrow$}
\caption{Cross section of the model system taken at $y=0$. 
The red areas correspond to electrodes that are spaced by $D=2.4\mu$m. The lower electrode
contains a circular tube extension with the inner radius $R_1=400$nm
and the outer one $R_2=500$nm. The dashed line is the symmetry axis
of the tube. The lower plate electrode is submerged by a liquid
helium layer of depth $H=1.2\mu$m. The distance from the top of the tube electrode to
the liquid surface is $d$. $\varphi$ stands for the electron wave function
localized above the liquid surface, close to the tube. 
}
\label{sc}
\end{figure}

Evaluation of the effective potential felt by an electron bound above the tube gate requires
solution of the generalized Poisson equation
\begin{equation}
\nabla\cdot\left(\varepsilon({\bf r})  \nabla V({\bf r})\right) = -\frac{\rho({\bf r})}{\epsilon_0},
\end{equation}
with $\rho({\bf r})=-e\delta({\bf r}-{\bf r}_e)$, where $\delta$ is the Dirac delta, and ${\bf r}_e$
is the electron position. The dielectric constant of liquid $^3$He is taken $\epsilon=1.0572$   \cite{g74}.
For the electron localized off the axis of the tube electrode, the $V$ potential does not possess 
the rotational symmetry and the problem requires a solution in a 3D computational box. 
The rectangular box is $4.8\mu$m wide in  $x$ and $y$ coordinates 
with height of $D$m (see Fig. 1) in the $z$ direction (we use $D=2.4\mu$m unless stated otherwise).
At the surface of the electrodes, the potential $V_1$ or $V_2$ is taken as the Dirichlet boundary condition
for the potential.  In the experiment \cite{kawakami21} the electric field of the order 
of up to $10$V/cm is applied to additionally press the electrons to the helium surface \cite{g74,yunusowa,kawakami21}. The field of this range does not qualitatively change the results
so we keep $V_1=V_2$ and use $V_1$ as the reference potential. 
For the low-energy electron states confined above the tube, we need to determine the potential in the region
of up to $\simeq 0.75\mu$m from the axis of the tube. At the sides of the computational box that are parallel to the $z$ axis, we apply the boundary condition of a vanishing component of the electric field ($\nabla V|_{\bf n}=0$) normal to the side of the computational box.

The Poisson equation is solved with the 
finite-element method \cite{fem} using parabolic Lagrange-type shape function in each element (see the Appendix).
The method allows us to take a zoom of the potential within the area occupied by electron states.
In the finite element method, the potential is expressed in the basis of shape functions $\Psi_l$,
\begin{equation} V({\bf r})=\sum_l c_l \Psi_l({\bf r}) \end{equation} associated with a single node each, so that 
$\Psi_l({\bf r_k})=\delta_{lk}$.
 
The Poisson equation is solved in the space spanned by the basis functions.
We substitute the expansion (2) to Eq. (1) and next project the result on a shape function $\Psi_k$ with integration over the computational box $\Omega$,
\begin{equation}
\int_\Omega \sum_{l} c_l \Psi_k({\bf r}) \nabla\cdot \left(\varepsilon({\bf r})  \nabla \Psi_l({\bf r}) \right)  d{\bf r}= -\int_\Omega \Psi_k({\bf r}) \frac{\rho({\bf r})}{\epsilon_0}d{\bf r} .
\end{equation}
Integrating by parts we get rid of the derivative over the dielectric constant,
\begin{eqnarray}
&& \sum_{l}c_l\int_\Omega \nabla  \Psi_k({\bf r}) \cdot \left(\varepsilon({\bf r})   \nabla \Psi_l({\bf r}) \right)d{\bf r} \nonumber\\
&&- \sum_{l}c_l\int_S   \Psi_k({\bf r})\varepsilon({\bf r})   \nabla \Psi_l({\bf r}) \cdot {\bf n}
d{\bf S} \nonumber\\
&&= \int_\Omega \Psi_k({\bf r})\frac{\rho({\bf r})}{\epsilon_0}d{\bf r},\label{seq}
\end{eqnarray}
where $S$ is the surface of the computational box, and ${\bf n}$ is a unit vector normal to the surface.
Equation (\ref{seq}) is a linear system of equations for $c_l$. Implementation of the boundary conditions for the nodes at the surface is explained in the Appendix.

\begin{figure*}[htbp]
\centering
\begin{tabular}{lllll}
\includegraphics[height=0.25\textwidth]{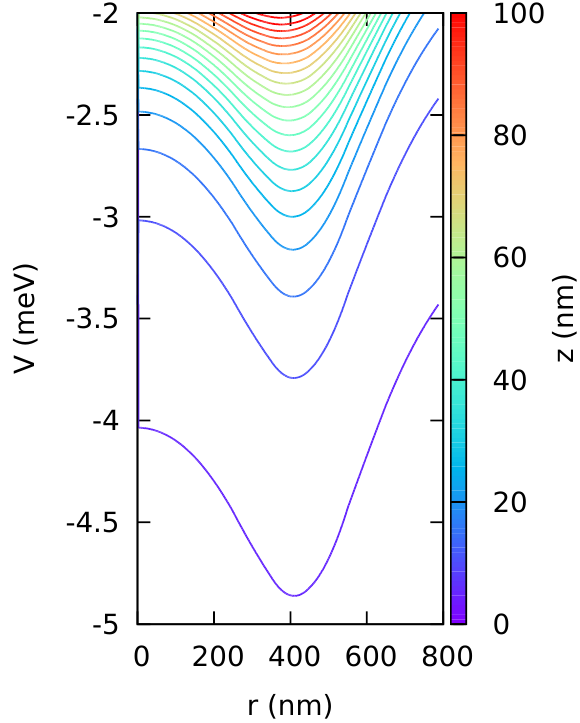}  \put(-77,43) {(a)}
&
\includegraphics[height=0.25\textwidth]{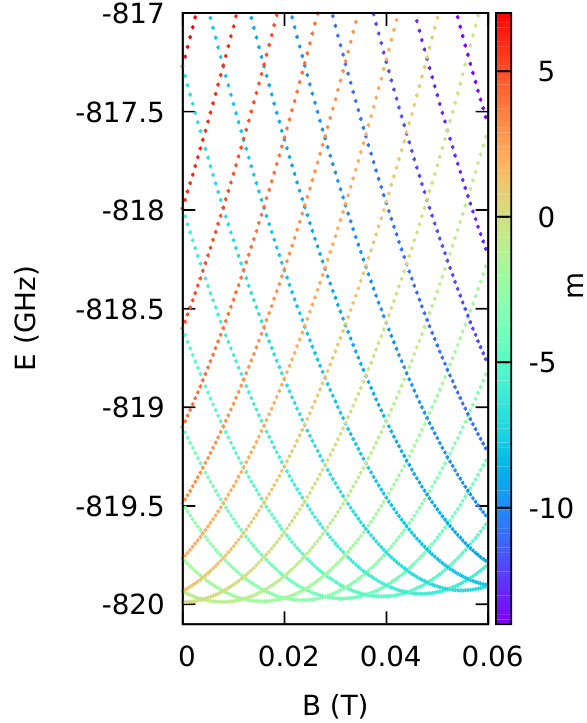} \put(-70,43) {(b)} &
 \includegraphics[height=0.25\textwidth]{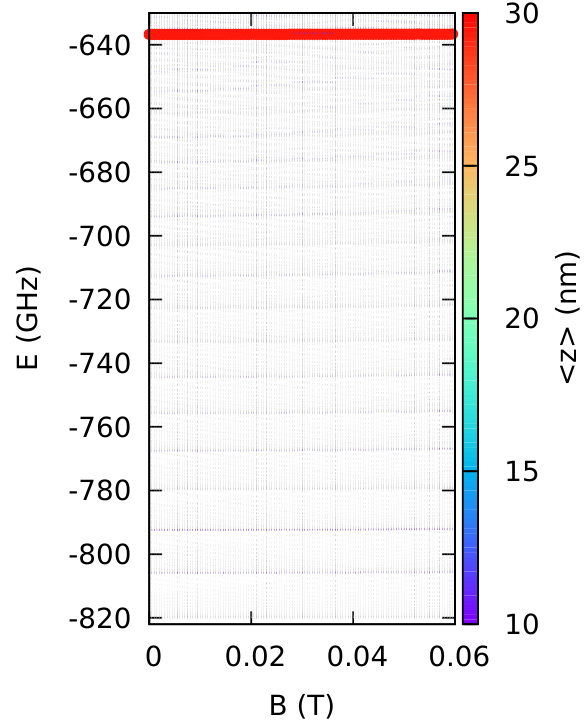} \put(-77,43) {(c)}\put(-39,18){\tiny \color{black}(0,0)}\put(-39,28) {\tiny \color{blue}(1,0)} \put(-39,35){\tiny \color{blue}(2,0)}  \put(-39,42){\tiny \color{blue}(3,0)}\put(-39,50){\tiny \color{blue}(4,0)} \put(-39,58){\tiny \color{blue}(5,0)} \put(-39,115){\tiny \color{red}(0,1)} & 
\includegraphics[height=0.25\textwidth]{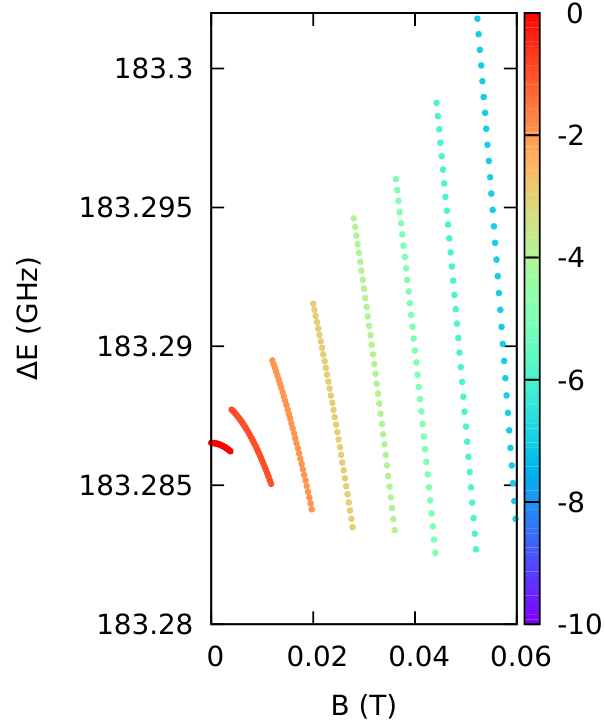} \put(-51,23) {(d)}
\includegraphics[height=0.25\textwidth]{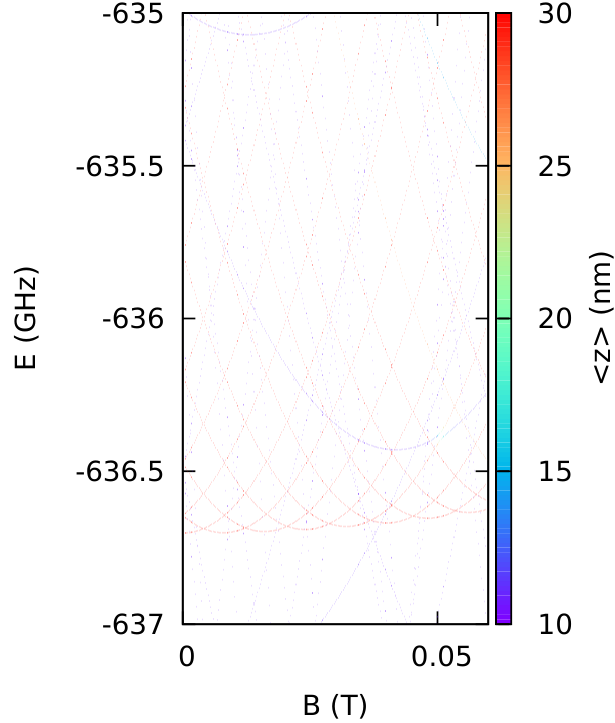} \put(-66,23) {(e)}
\end{tabular}
\caption{The results for the tube electrode at $d=200$ nm beneath the surface of liquid helium.
(a) The confinement potential $W$ for the electron in vacuum. The color of the lines
indicates the distance $z$ from the liquid helium surface. The lowest line is taken at $z=5$ nm above the surface,
and the higher lines are plotted at spacings of 5 nm.
(b) The lowest-energy part of the spectrum in a vertical magnetic field. The color of the lines
indicates the angular momentum quantum number $m$. The ground-state at $B=0$ is $m=0$, and changes by $-1$ with
each crossing of energy levels as $B$ is increased.
(c) The energy spectrum for a wider range of energy. The energy levels for $m=-22,-21,\dots,11$ are marked
by gray lines. The lines marked by colors indicates the energy levels that are dipole coupled to the ground state.
The lines width is proportional to the absolute value of the dipole matrix element $d_{if}=\langle\phi_i|z|\phi_f\rangle$. 
$(n_r,n_z)$ indicates the subbands of the energy levels, with $n_r$ and $n_z$ standing for the number
of wave function zeroes in the radial and vertical directions respectively.
The color of the lines shows the average $z$ position of the corresponding energy level wave function. 
(d) The absorption spectrum calculated as the energy difference between the excited state energy levels
and the ground-state. The size of the dots is proportional to the absolute value of the dipole matrix element. 
The presented range corresponds to $(0,0)\rightarrow(0,1)$ transition (first to second Rydberg level).
(e) The part of the energy spectrum corresponding to the (0,1) subband. The color of the lines shows the average $z$ position. }
\label{wf}
\end{figure*}

\begin{figure}[htbp]
\centering
\begin{tabular}{l}
\includegraphics[height=0.12\textwidth]{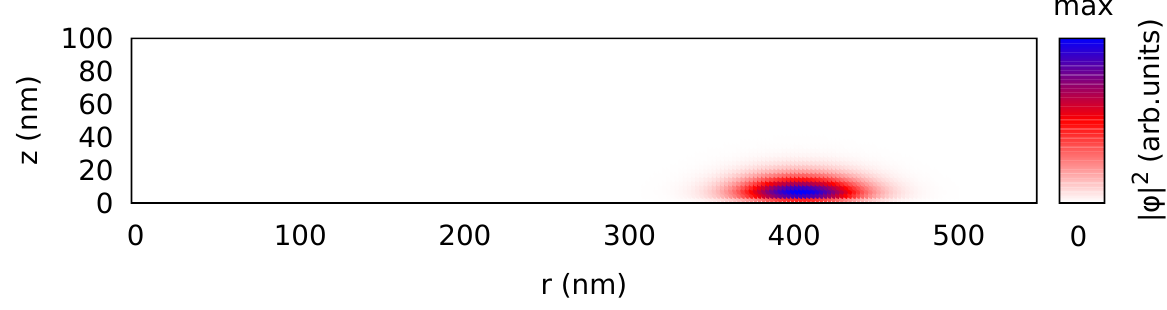} \put(-40,45){(a)}   
\put(-180,45){\tiny \color{black}$(n_r,n_z)=(0,0)$}\\
\includegraphics[height=0.12\textwidth]{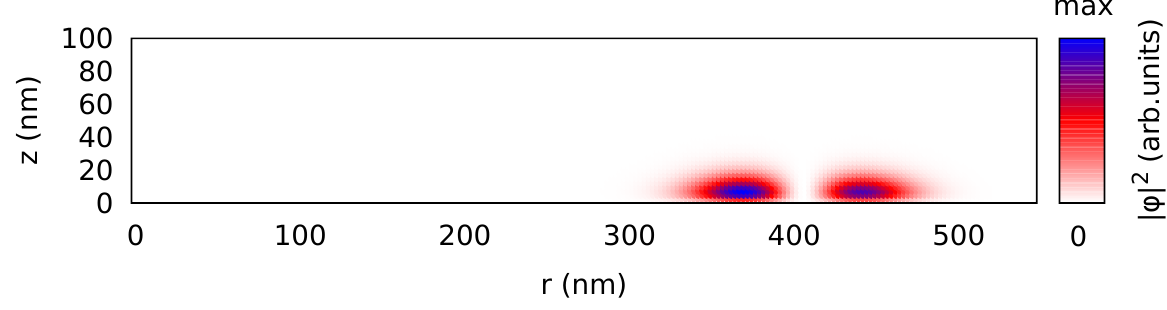} \put(-40,45){(b)}  \put(-180,45){\tiny \color{black}$(1,0)$}\\
\includegraphics[height=0.12\textwidth]{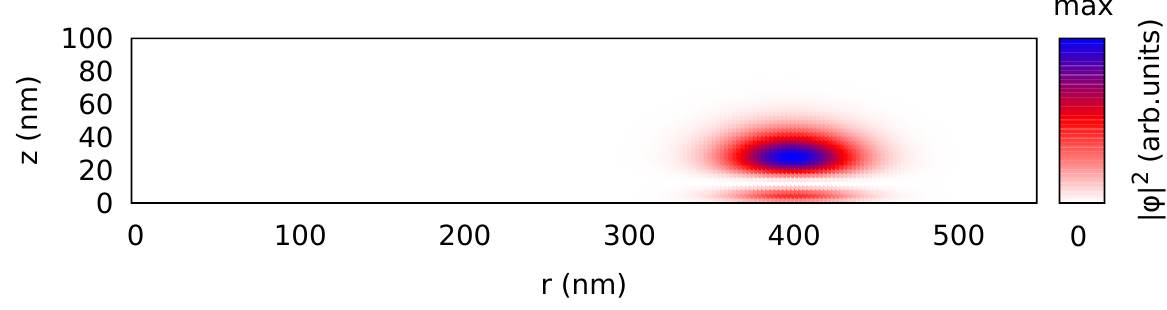} \put(-40,45){(c)}  \put(-180,45){\tiny \color{black}$(0,1)$}\\
\end{tabular}
\caption{Probability density for $m=-1$ states at $B=0$ for $d=200$nm. (a),(b) and (c) correspond
to $(n_r,n_z)=(0,0)$, $(1,0)$ and $(0,1)$ states.
 }
\label{wv10}
\end{figure}

\begin{figure*}[htbp]
\centering
\begin{tabular}{lllll}
\includegraphics[height=0.25\textwidth]{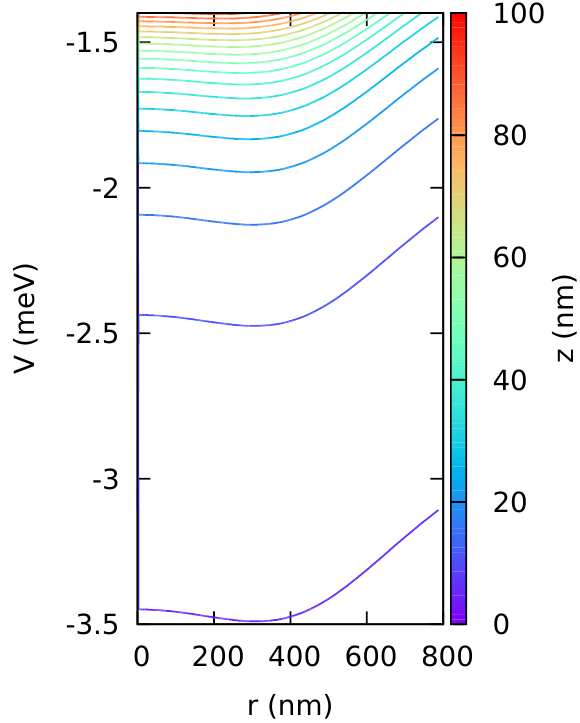} \put(-38,20) {(a)} &
\includegraphics[height=0.25\textwidth]{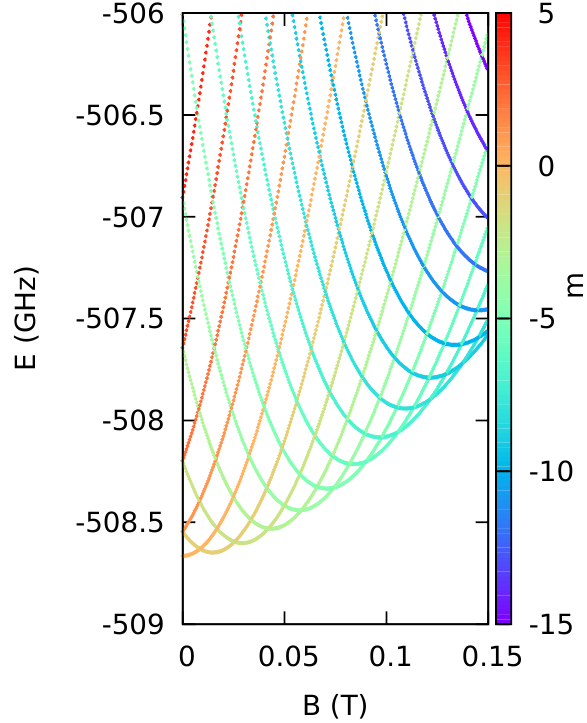} \put(-30,20) {(b)} &
 \includegraphics[height=0.25\textwidth]{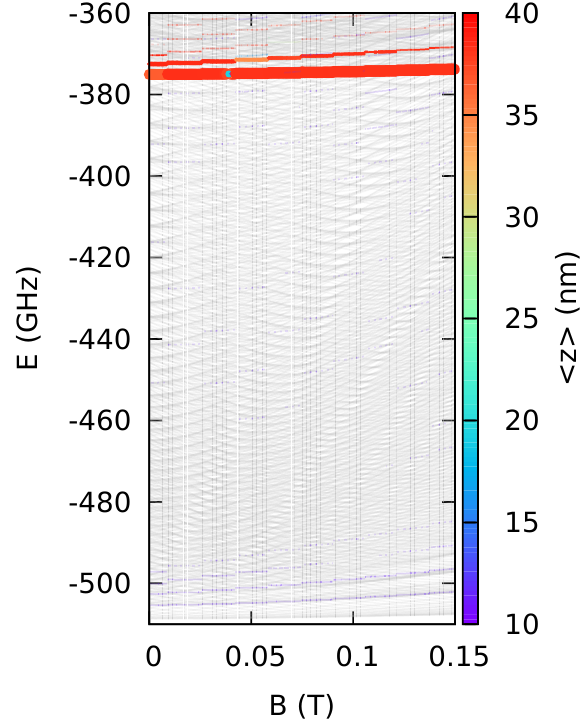} \put(-74,20) {(c)}
  \put(-39,18)
   {\tiny \color{blue}(1,0)} \put(-39,24){\tiny \color{blue}(2,0)} 
    \put(-53,26)
    {\tiny \color{blue}(3,0)}\put(-39,34){\tiny \color{blue}(4,0)} 
\put(-42,110){\tiny \color{red}(0,1)} 
 \put(-42,118){\tiny \color{red}(0,2)} & 
\includegraphics[height=0.25\textwidth]{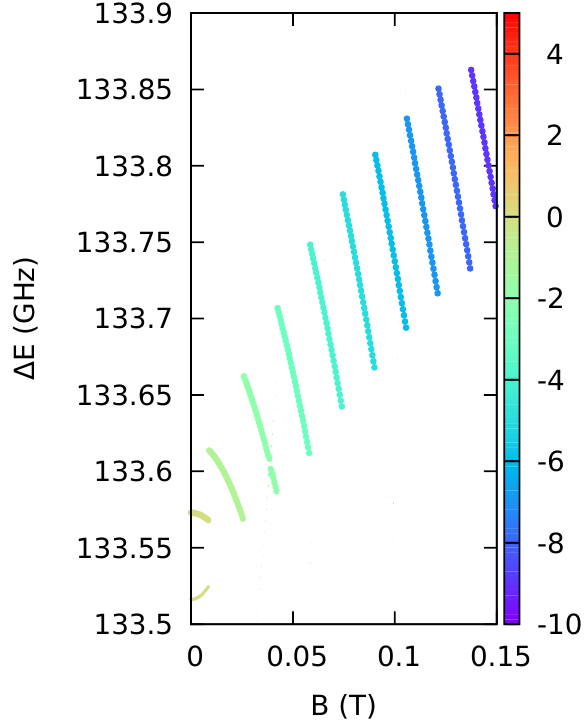} \put(-28,20) {(d)} 
\includegraphics[height=0.25\textwidth]{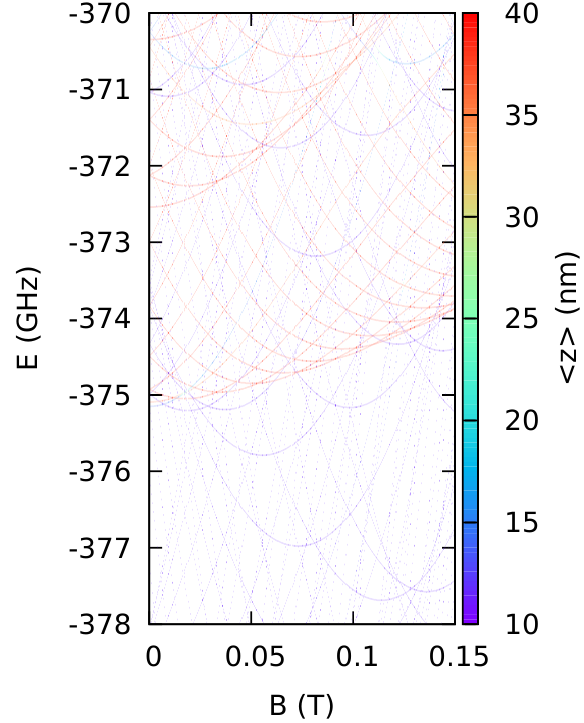} \put(-34,20) {(e)}\\
\end{tabular}
\caption{Same as Fig. 2, only for $d=400$nm.}
\label{400}
\end{figure*}

\begin{figure}[htbp]
\centering
\begin{tabular}{l}
\includegraphics[height=0.12\textwidth]{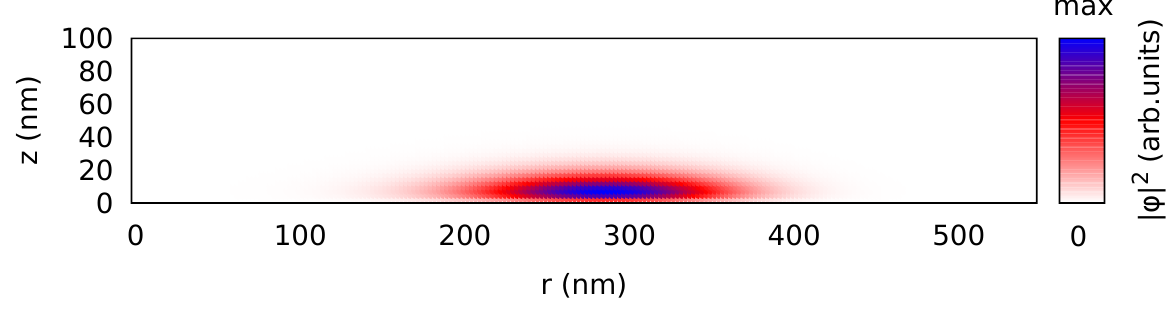}  \put(-40,40){(a)} \put(-120,50){\tiny \color{black}$(n_r,n_z)=(0,0)$}\\
\includegraphics[height=0.12\textwidth]{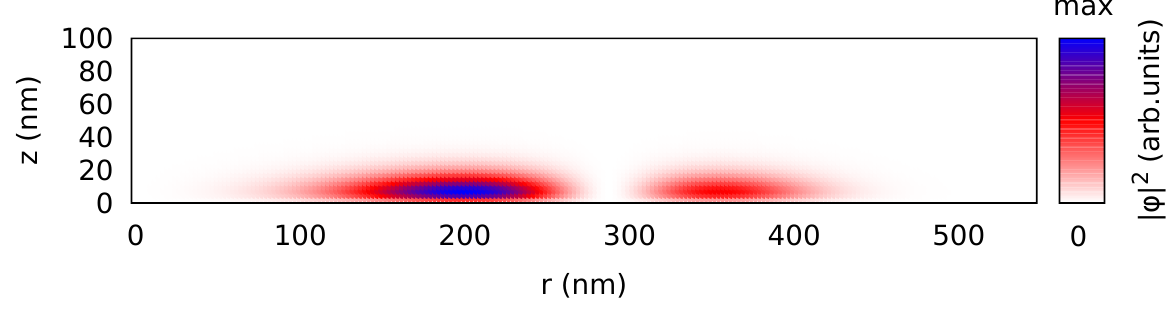} \put(-40,40){(b)}\put(-120,50){\tiny \color{black}$(1,0)$}\\
\includegraphics[height=0.12\textwidth]{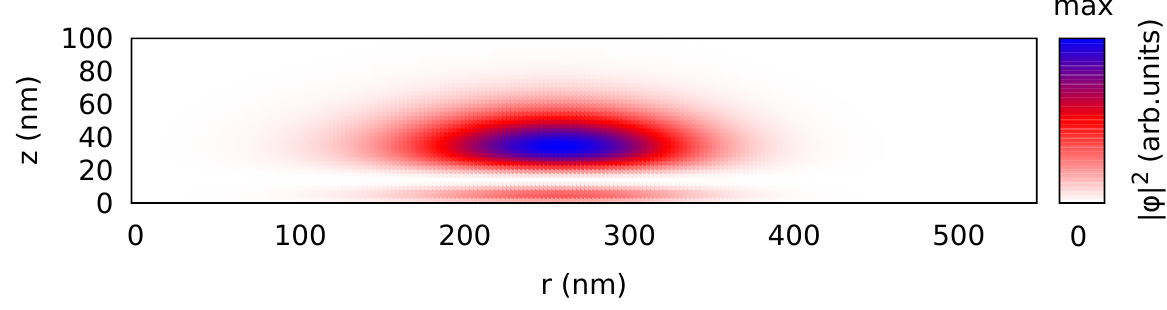} \put(-40,40){(c)}\put(-120,50){\tiny \color{black}$(0,1)$}\\
\end{tabular}
\caption{Same as Fig. 3, only for $d=400$nm.}
\label{wv400}
\end{figure}

We solve the Poisson equations for the electron position ${\bf r_e}$ scanned over the finite element nodes above the liquid helium.
The potential $V(\bf r)$ calculated in this way contains both the contributions from the point charge of the electron floating
above the helium surface and the charges induced in liquid helium and on metal electrodes. 
To determine the effective potential acting on the electron near the helium surface, we need to
eliminate the electron self-interaction included in $V$ \cite{lis1}.
For this purpose, we calculate potential $V_e$ for the electron above the helium surface but with  all the metal electrodes removed. 
For $V_e$ we apply the Dirichlet boundary conditions at all the sides of the computational box using the  potential for the point charge near the contact of two dielectrics 
 i.e. $V_e=\frac{1}{4\pi\epsilon_0}\left(\frac{e}{R_e}+\frac{q'}{R_i}\right)$ for $z$ above
 the helium layer, where $R_e=|{\bf r}-{\bf r}_e|$ is the distance from the electron position and $R_i=|{\bf r}-{\bf r}'_e|$ the distance from its image placed at ${\bf r}'_e=(x_e,y_e,H-(z_e-H))$  with the image charge $q'=\frac{1-\varepsilon}{1+\varepsilon}e$. At the sides of the computational box below the helium surface we take at the boundary the potential $V_e=\frac{1}{4\pi\varepsilon\epsilon_0}\frac{q''}{R_e}$, where
 $q''=\frac{2\varepsilon}{1+\varepsilon}e$.
 
With $V$ and $V_e$ we calculate an effective potential $V_1=V-V_e$, in which the electron self-interaction is removed.
 Since in the calculation for both $V$ and $V_e$ we account for the presence of the helium surface, in the evaluation of $V_1$ the image charge potential induced on the liquid surface is canceled. In the Schroedinger equation that we solve for the confined states we reintroduce 
the potential of the image charges induced in helium  in its analytical form \cite{g74}, $W_s(z_e)=-\frac{Ze^2}{|z_e-H|}$, with $Z=\frac{\varepsilon-1}{4(1+\varepsilon)}$. $W_s$ potential diverges for $z\rightarrow H$, hence it is more readily introduced  in its analytical form to the Schroedinger equation than evaluated in the polynomial shape functions. 
The confinement potential energy to be used in the Schroedinger equation is then set as $W({\bf r_e})=-e\left(V{(\bf r_e})-V_e{(\bf r_e})\right)+W_s({\bf r_e})$. We assume that the liquid helium is impenetrable
 for the low-energy electrons of the vacuum side and take the infinite value of $W$ below the surface.

 Although for a general electron position ${\bf r_e}$, the potential $V({\bf r})$ is not rotationally symmetric, 
 the effective potential $W({\bf r_e})$ has this symmetry.
 We solve the Schroedinger equation in cylindrical coordinates for a given angular momentum 
 quantum number $m$, $\varphi_m=\varphi(r,z)\exp(im\theta)$, with azimuthal angle $\theta$. 
With the symmetric gauge ${\bf A}=(-By/2,Bx/2,0)$ for vertical magnetic field ${\bf B}=(0,0,B)$ 
the Hamiltonian reads,
 \begin{eqnarray} H&=&-\frac{\hbar^2}{2m_0}\left[\frac{\partial^2}{\partial z^2}
 +\frac{\partial^2}{\partial r^2}+\frac{1}{r}\frac{\partial}{\partial r}-\frac{m^2}{r^2}\right] \\ \nonumber
&& +\frac{m\hbar \omega_c}{2}+\frac{m_0}{8} \omega_c^2 r^2+W(r,z),
 \end{eqnarray}
 with $\omega_c=\frac{eB}{m_0}$.
 We use the imaginary time method \cite{imaginary} to determine the eigenstates of the finite difference version
 of the Hamiltonian with the mesh spacing of 2.5nm in $r$ and $z$ coordinates.
 The region occupied by electrons in low-energy states is much smaller than the computational box
 for the Poisson equation and a finer mesh is required to describe the wave functions.
The potential spanned by the shape functions $W(r,z)$ can be evaluated on a finer grid.
\begin{figure*}[htbp]
\centering
\begin{tabular}{llll}
  \includegraphics[height=0.28\textwidth]{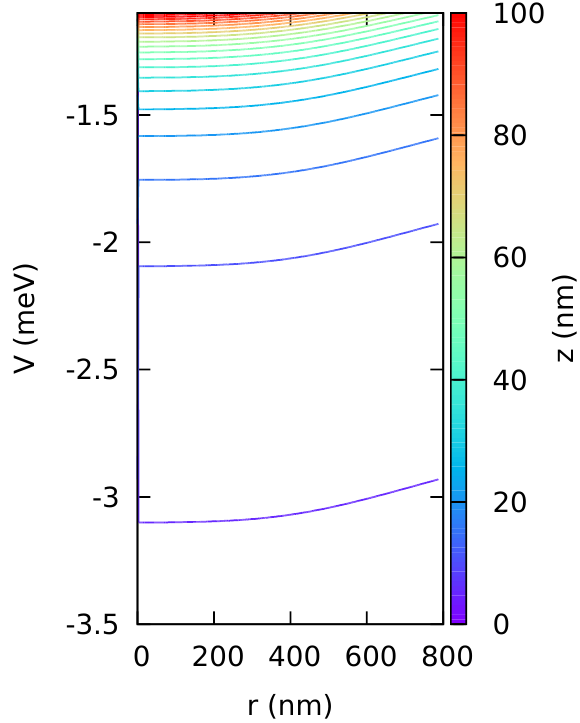} \put(-40,23) {(a)} &
\includegraphics[height=0.28\textwidth]{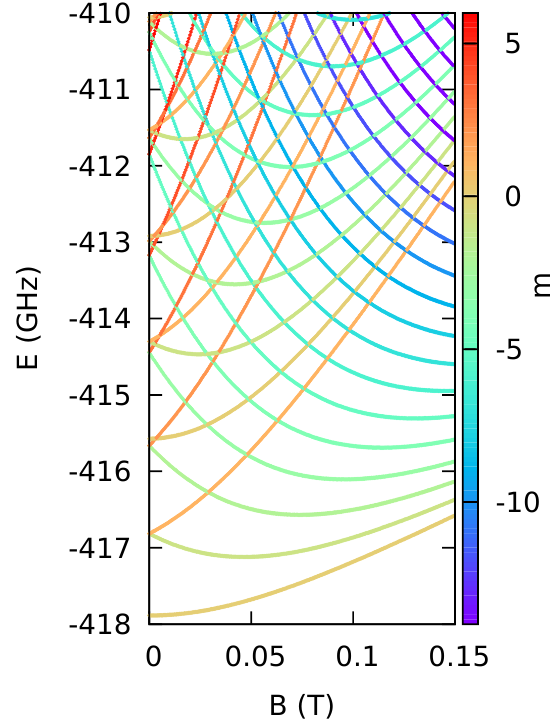} \put(-31,23) {(b)} & 
\includegraphics[height=0.285\textwidth]{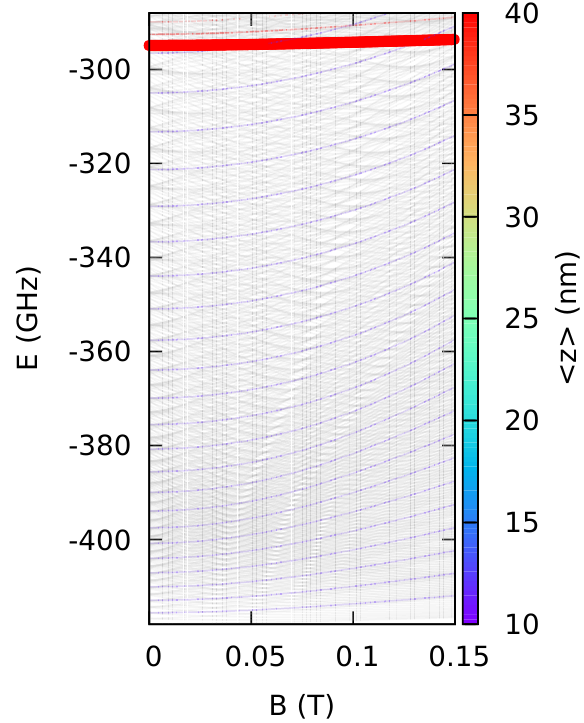} \put(-38,23) {(c)} &
\includegraphics[height=0.28\textwidth]{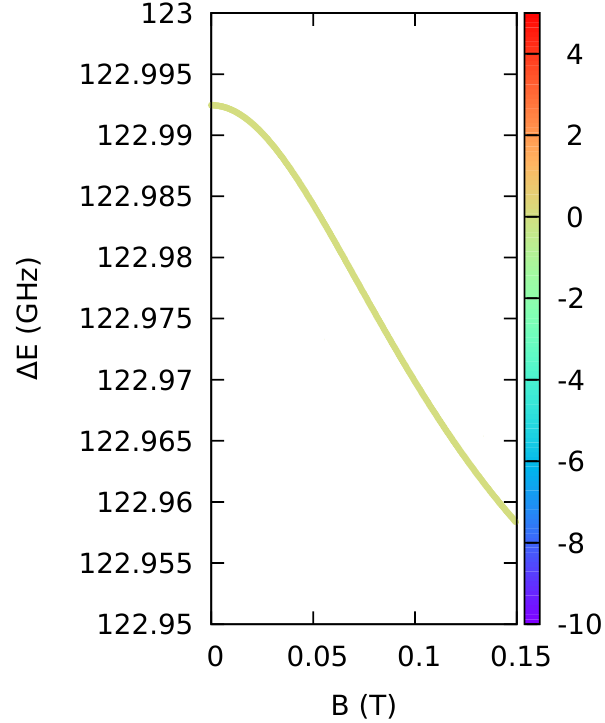} \put(-30,23) {(d)} \\ 
\end{tabular}
\caption{Same as Fig. 2 only for $d=600$ nm.}
\label{600}
\end{figure*}

\section{Results}
\subsection{Shallow tube ($d=200$nm)}

For the presentation of the results, we fix $z=0$ at the surface of the liquid helium.
The electron confinement potential depends qualitatively on the depth
at which the tube electrode is located beneath the surface. 
For $d=200$nm the potential has a pronounced local maximum near $r=0$ [Fig. 2(a)] and
the minimum is located near $r=400$nm.
The minimum gets shallower and shifts to lower values of $r$ at a larger distance from the surface.
The changing profile of the lateral potential with $z$ indicates a non-separability of the potential,
in contrast to a separable potential found for a submerged rod-like electrode \cite{Dy}.

The energy spectrum near the ground state is shown in Fig. 2(b). The ground-state quantum number $m$
 changes as for the semiconductor quantum rings \cite{qr,fuhrer}. 
For a strictly one-dimensional ring of radius $R_0$, the energy spectrum is given by
$E_m=\frac{\hbar^2}{2m_0R_0^2}\left(\frac{\Phi}{\Phi_0}+m\right)^2$, with flux quantum $\Phi_0=\frac{e}{h}$,
and the flux threading the quantum ring $\Phi=B\pi R_0^2$. 
In Fig. 2(b) we observe the changes of the ground-state angular momentum
in $B$ with the spacing of $\Delta B\simeq 8$ mT.
The radius  corresponds to the magnetic field flux through a one-dimensional ring of $R\simeq 430$nm.

The spectrum in a wider range is displayed in Fig. 2(c). The energy levels for $m=-22,-21,\dots 11$ are marked in gray. Additionally, we marked with colors those excited energy levels
that are dipole coupled to the ground state. The width of the color lines is proportional to the
dipole matrix element $d_{if}=|\langle \varphi_{i} |z| \varphi_{f}\rangle|$, where $\varphi_{i}$ and $\varphi_{f}$ stand for the states participating in the transition from the initial (ground) and final (excited) state wave functions.
 The color of the lines  indicates the
average $z$ for the excited states. In the spectrum we see a set of thin blue
lines for states  localized at $z\simeq 12$nm. The excited states to which the transition is allowed correspond  to the same $m$ as the ground state but correspond to excitations within the plane of confinement. 
For separable
potential the wave function can be put in form $\varphi_{n_r,n_z}(r,z)=\varphi^r_{n_r}(r)\varphi^z_{n_z}(z)$,
and the dipole matrix element for transition from the ground state would be 0 for the final states with $n_r\neq 0$ due to the in-plane orthogonality. However, the potential $W$ is not a simple sum of the vertical and lateral components [Fig. 2(a)] and the wave function is non-separable, which allows for a small but nonzero dipole moment. 
Although the potential is non-separable, one can attribute to states the number of wave function zeroes along the $r$ and $z$ directions, $n_r$ and $n_z$, where $n_z+1$ numbers the Rydberg subbands. 
At the top of Fig. 2(c) we see a thick red line that corresponds to the second Rydberg subband  with a vertical excitation (0,1). The probability densities  for $m=-1$ and $(n_r,n_z)$ equal to (0,0), (1,0), and (0,1) are plotted
in Fig. \ref{wv10}(a),(b), and (c), respectively.
The spectrum of the second Rydberg subband (0,1) [Fig. 2(d)] has a similar quantum-ring pattern
as the ground-state subband (0,0). The period is almost the same as for the (0,0) subband of Fig. 2(b),
since these states are localized at similar $\langle r \rangle$ (cf. Figs. 3(a) and 3(c)). 
The absorption spectrum $\Delta E$ for the transitions from the ground state is depicted in Fig. 2(e).
The spectrum indicates that the main transition line from the lowest to the second Rydberg subband is nearly independent
of the external magnetic field. The $m$ changes in the ground-state (0,0) subband and in the (0,1) subband are nearly perfectly synchronized
on the $B$ scale. As a consequence, the Aharonov-Bohm angular momentum transitions produce only weak symptoms
on the energy spectrum of the microwave absorption with fine discontinuities of the absorption line
that jump by less than $0.02$ GHz.

\subsection{Discontinuous transition spectrum ($d=400$ nm)}

The period of  angular momentum transitions can be made unequal in the lowest (0,0) and second (0,1) Rydberg subbands for the tube electrode submerged deeper in helium. The confinement potential for the tube electrode at a depth of $d=400$nm beneath the surface is displayed in Fig. \ref{400}(a). 
Compared to Fig. 2(a), the potential minimum is shifted towards the axis of the system
and the position of the minimum changes as one rises above the surface to eventually disappear near $z=100$nm. 
Compared to Fig. 2 the ground-state angular momentum transitions occur with an increased spacing on the magnetic field scale
due to reduction of the confinement ring radius [see Fig. 3(b)]. The energy level that is coupled to the
(0,0) ground-state is discontinuous  as a function of $B$ (Fig. \ref{400}(c)). In the
second Rydberg (0,1) subband, the angular momentum changes with a slower rate:  the wave functions -- are localized higher above the surface and are localized closer to the axis than in the lowest (0,0) subband [cf. Fig.\ref{wv400}(a,c)].
 As a consequence, the angular momentum transitions in the first and second Rydberg subbands no longer occur at the same values of $B$ and the absorption spectrum (Fig. \ref{400}(e)) contains
more pronounced discontinuities with jumps of the line of the order of 0.1 GHz. 

\subsection{Deep tube ($d=600$ nm)}

For the tube $d=600$nm beneath the helium surface, the $W$ potential loses its local maximum  at the axis [Fig. \ref{600}(a)]. The energy spectrum [Fig. \ref{600}(b)] is nearly identical with the Fock-Darwin one \cite{fd},
but with lifted degeneracy of the excited energy levels of non-equal $|m$| due to the deviation of the confinement
potential from parabolicity. The main absorption line  [Fig. \ref{600}(c,d)]  is nearly independent of the external magnetic field. 

\subsection{Geometry effects}
\begin{figure}[htbp]
\centering
  \includegraphics[height=0.28\textwidth]{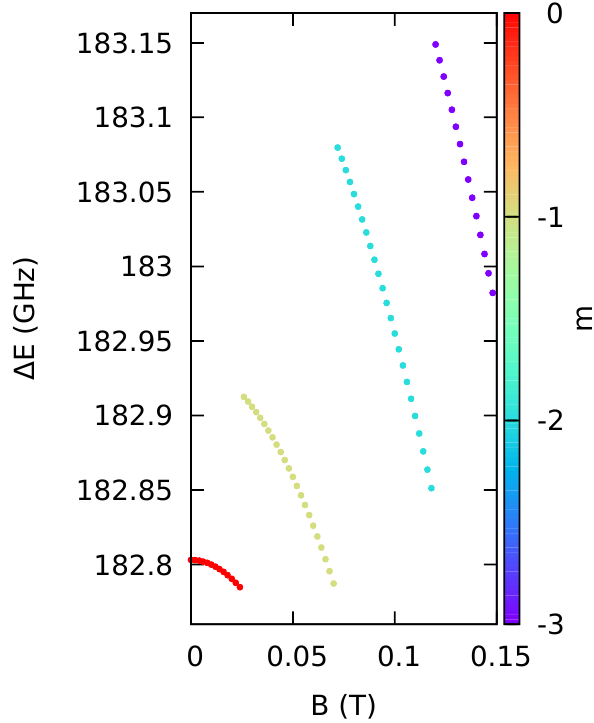} \put(-40,23) {(a)}
  \includegraphics[height=0.28\textwidth]{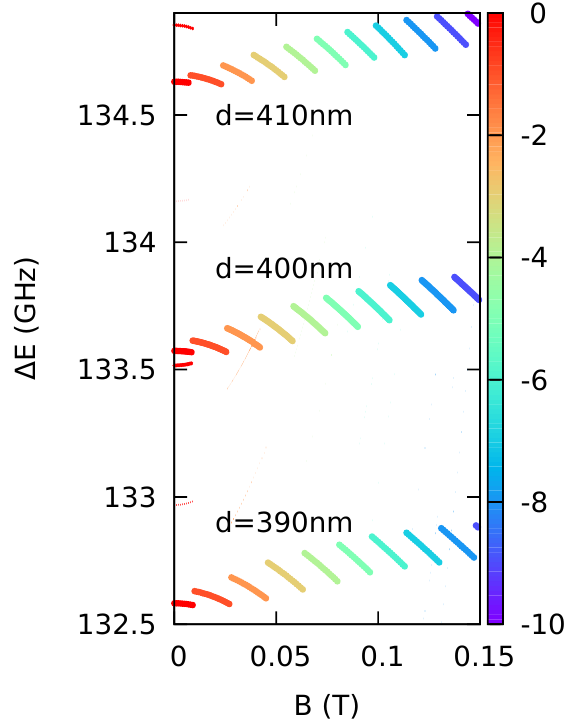} \put(-40,23) {(b)}
  \caption{Absorption spectra of the transition from the lowest to the second Rydberg level. The colors
show the magnetic quantum number $m$. In (a) a thin tube with radii $R_1=300$ nm and $R_2=400$ nm  covered with helium layer of depth $d=200$ nm is taken. In (b) 
  $R_1=400$nm and $R_2=500$nm are applied as elsewhere in this paper and the values of $d$ are varied.}
\label{34}
\end{figure}

The discontinuities of the absorption spectrum can be made larger for tuned
geometry. The spectrum with the inner and outer radii reduced to $R_1=300$ nm and $R_2=400$ nm is displayed in Fig. \ref{34}(a),
where we use a thinner layer of helium,
 $d=200$ nm instead of $d=400$ nm.
For the tube of reduced radius, the confinement potential changes faster with $z$ for these parameters which results in the increased discontinuities
in the spectrum, e.g. the ground-state transition from $m=-2$ to $m=-3$ involves
a jump of 0.3 GHz instead of 0.13 GHz as in Fig. \ref{400}(d).

For the rest of the paper, we return to $R_1=400$ nm, $R_2=500$ nm and $d=400$ nm.
The sensitivity of the spectrum with respect to the exact  width of He layer 
above the tube near the work point of $d=400$ nm can be studied in Fig. \ref{34}(b).
The central set of lines in the spectrum is the one of Fig. \ref{400}(d) with $d=400$ nm. 
The lower set corresponds to $d=390$ nm and the upper one to $d=410$ nm.
The 10 nm change of the liquid layer depth induces a shift of the absorption line by 1.1 GHz.
For lower values of the width of the helium layer $d$ above the tube, the effective confinement radius gets larger,
hence a slight modification of the oscillation period.


In experimental conditions the distance between the plates of the electrodes needs to be larger than the liquid helium capillary length, i.e., not less than 1.5 mm.
Our computational model is much smaller.  
The length of the tube cannot be increased much above several $\mu$m due to the limitations of the focused ion beam methods
that can define such an electrode. Therefore, the upper plate needs to be further away from the helium surface. 
To study this size effect, we kept $H$ and $d$ unchanged and increased $D$ (see Fig. 1) 
from 2.4 $\mu$m (as in all the results above) to 12.4 $\mu$m and $22.4$ $\mu$m. The results are given in Fig. 8.
The image charge at the upper capacitor plate produces an electric field that pulls the electron up from the surface.
This field is reduced for larger $D$. 
Fig. 8 shows that as $D$ grows, the energy position of the main absorption line is going up on the energy scale
in consistence with the study of the effect of the vertical electric field on the transition between the ground and the  excited Rydberg states \cite{g74,yunusowa,kawakami21}. At a larger distance between the top plate 
and the helium surface, the discontinuities of the absorption lines are preserved.
We do not expect qualitative changes of the character of the spectrum for $D$ increasing further.


\begin{figure}[htbp]
\centering
  \includegraphics[height=0.3\textwidth]{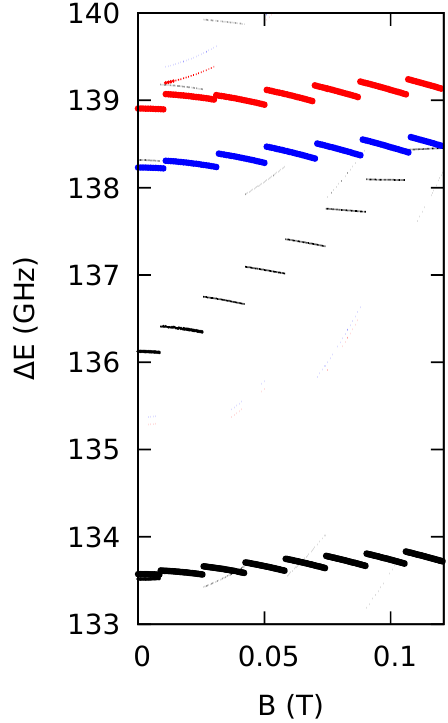} \put(-65,22){$D=2.4\mu$m}, 
  \color{blue} \put(-70,110){$12.4\mu$m}  \color{red} \put(-50,140){$22.4\mu$m} \color{black}
  \caption{Absorption spectra of the transition from the lowest to the second Rydberg level
  for $H=1200$ nm, $d=400$ nm, $R_1=400$ nm and $R_2=500$ nm.
  Results for $D=2.4\mu$m, $D=12.4\mu$m and $D=22.4\mu$m are plotted with black, blue and red lines, respectively.}
\label{size}
\end{figure}

\begin{figure*}[htbp]
\centering
\begin{tabular}{lll}
  \includegraphics[height=0.25\textwidth]{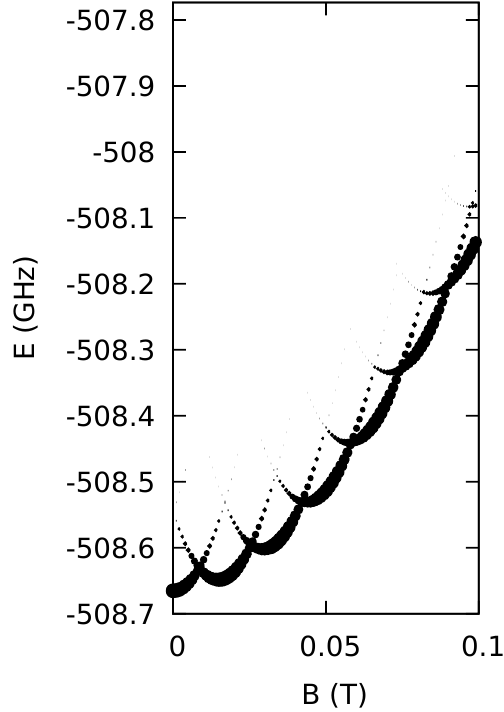} \put(-20,113) {(a)} \put(-50,113) {1mK}  &
  \includegraphics[height=0.252\textwidth]{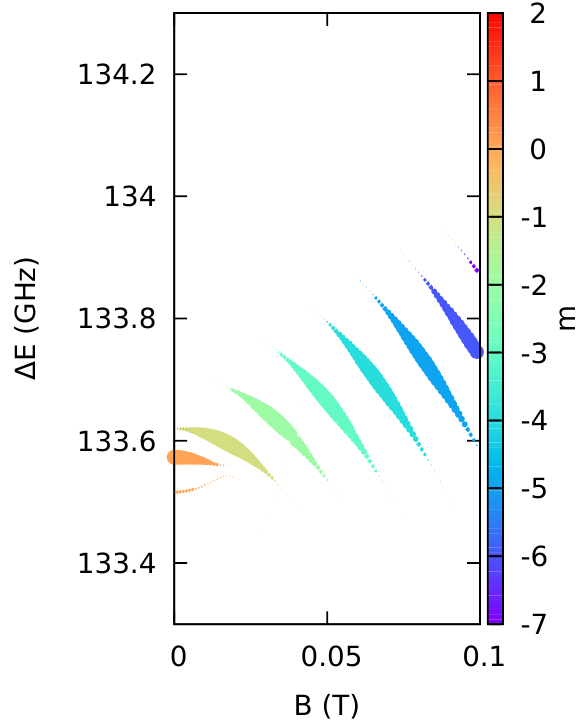} \put(-30,113) {(b)} &
    \includegraphics[height=0.252\textwidth]{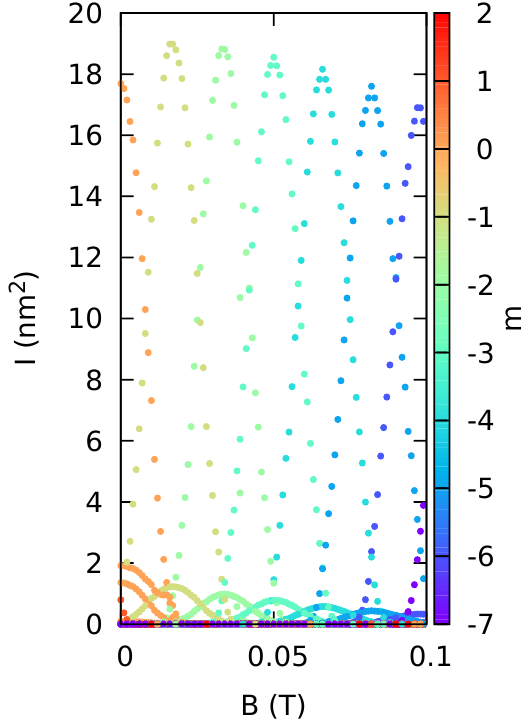} \put(-30,113) {(c)} \\
\includegraphics[height=0.25\textwidth]{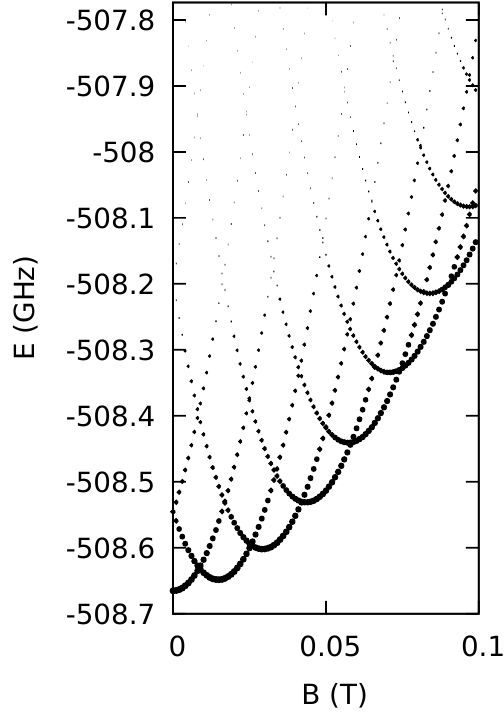} \put(-20,113) {(d)} \put(-50,113) {10mK}   & 
\includegraphics[height=0.252\textwidth]{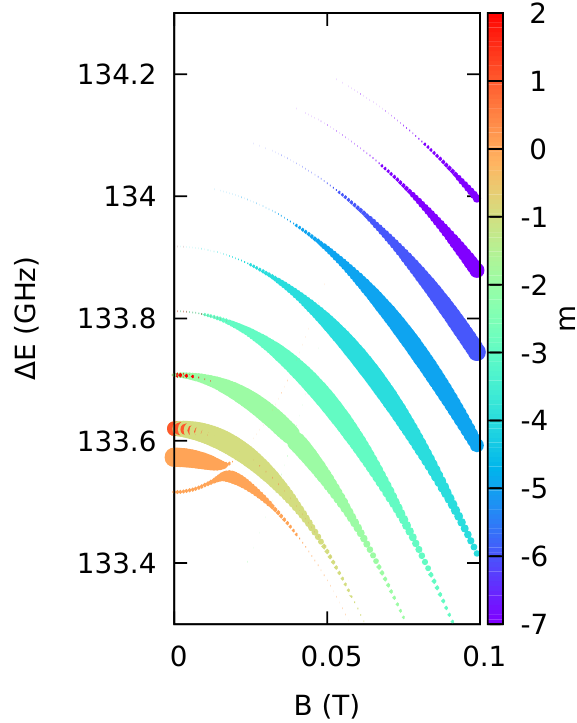} \put(-30,113) {(e)} & 
\includegraphics[height=0.252\textwidth]{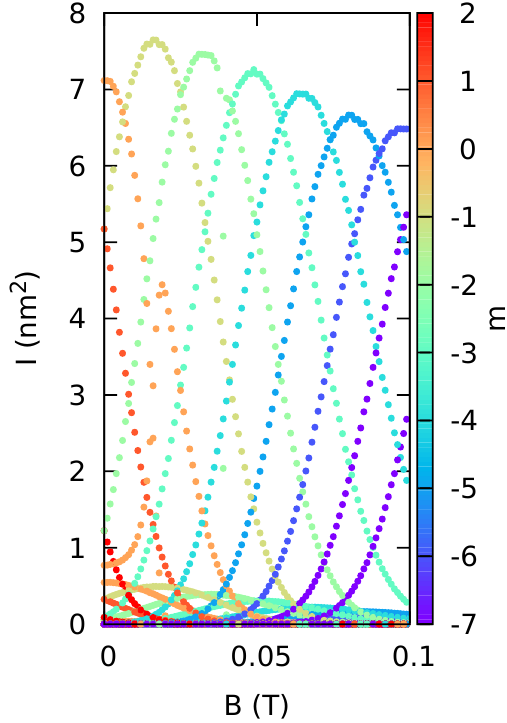} \put(-30,113) {(f)} \\
\includegraphics[height=0.25\textwidth]{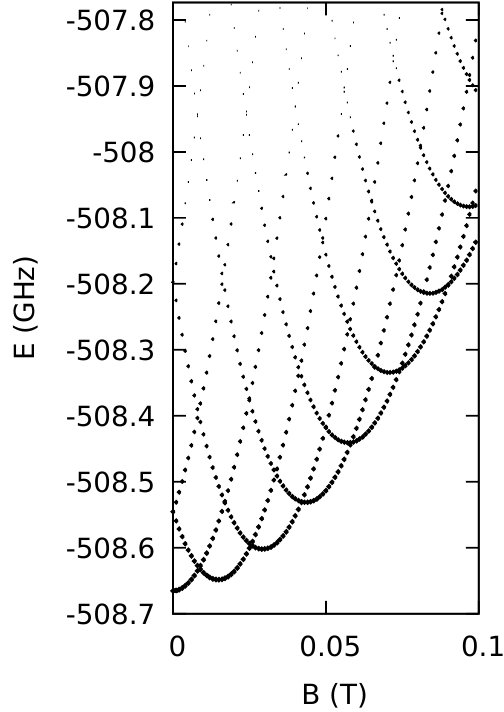} \put(-20,113) {(g)} \put(-50,113) {20mK}  &
\includegraphics[height=0.252\textwidth]{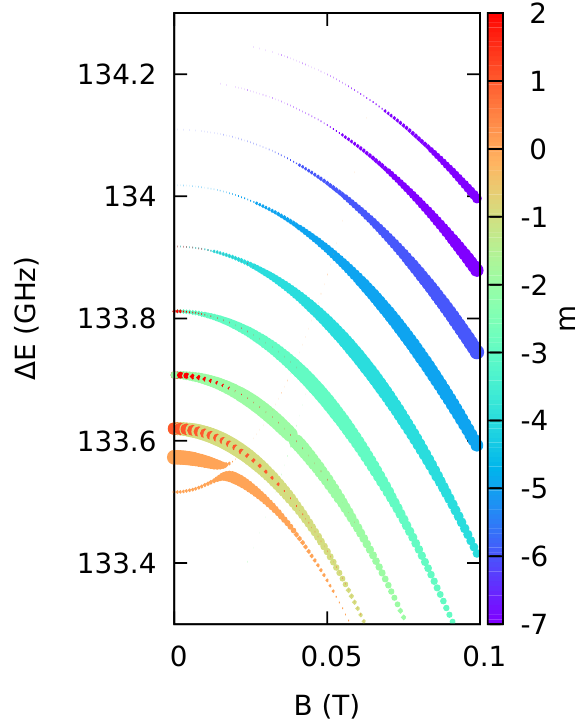} \put(-30,113) {(h)} &
\includegraphics[height=0.252\textwidth]{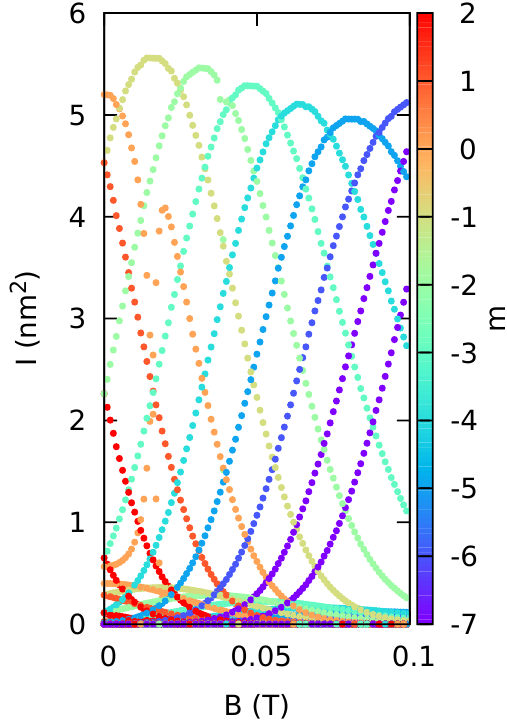} \put(-30,113) {(i)}

\end{tabular}
\caption{Results for the geometry of Fig. 4 in finite temperature. The first (a-c),
second (d-f) and third row (g-i) of plots correspond to $T=1$mK, 10 mK and 20 mK, respectively. In the left column
of plots (a,d,g) the dots mark the energy levels near the ground-state with their size proportional to the occupation of the energy level.
In the central column (b,e,h) the absorption spectrum is plotted with the size of the symbol
proportional to the product of the initial state occupation and the transition probability. The colors mark the angular momentum for the states participating in the  transition. The right column of plots shows the intensity of the line.}
\label{t}
\end{figure*}

\subsection{Finite temperature}
The absorption spectra presented above for zero temperature exhibit discontinuities
with the Aharonov-Bohm periodicity. The discontinuities result from the ground-state
angular momentum transitions driven by the external magnetic field. 
The energy spacings between states confined in the tube potential are small
and in the finite temperature several lines will be present at the same magnetic field.
We need to look for signatures of the absorption spectrum that will pertain
in finite $T$. 

The occupation probability of the energy level with energy $E_i$ 
is given by the Fermi-Dirac distribution $p_i=\frac{1}{\exp(\beta(E_i-\mu)+1}$,
where $\beta=\frac{1}{k_B T}$ and $\mu$ is the chemical potential that for a single-electron per tube solves the equation $\sum_i{p_i}=1$. 
The intensity of the absorption line for initial state $i$ and final state $f$
can be evaluated as $I_{if}=p_i |d_{if}|^2$.
The low-energy spectrum is presented in Fig. \ref{t}(a,d,g) with the dots
of the size that is proportional to $p_i$ for $T=1$mK, 10mK and 20 mK, respectively.
The simulated transition spectra are plotted in Fig. \ref{t}(b,e,h) with dots
of size that is  proportional to $I_{if}$. Already at $1$mK the magnetic field intervals 
for transitions to states of subsequent $m$ values overlap and at $10$mK the discontinuities in the absorption spectra are no longer present.
Nevertheless, the Aharonov-Bohm periodicity can still be extracted from the absorption spectrum
by observation of the intensities of separate lines [see Fig. \ref{t}(c,f,i)]. The intensities are maximal for the magnetic fields that appear in the middle of subsequent ground-state angular momentum transitions for which the energy spacing between the ground state and the excited states is the largest (see Fig. 4(b)). 
For $B$ that corresponds to the maximal intensity of line  $m$ the neighbor lines with $m\pm 1$ have equal
intensities, which can be useful as an additional feature for identification
of the confined Aharonov-Bohm effect.

\section{Discussion}

The results above indicate that the experimental detection of the Aharonov-Bohm ground-state transitions by microwave absorption 
requires tuning of the liquid helium level to conditions in which the radius of confinement in
the lowest and second Rydberg subbands is significantly different. Adjusting the liquid layer  above the tube electrode should be relatively straightforward. 
The discontinuity of the absorption energy line due to the ground-state transitions at $d=400$nm is of the order of 0.1 GHz [Fig. \ref{400}(d)] which is within range of the resolution of the microwave absorption measurements \cite{kawakami21,yunusowa,lambs}. 

The counterparts of the discussed angular momentum transitions in a single semiconductor quantum ring are resolved
by the transport spectroscopy for systems with attached contacts \cite{fuhrer}. 
In the experiment on electrons on a liquid helium surface, an array of tube electrodes could be considered. In self-assembled semiconductor quantum rings, the ground-state transitions induced by an external magnetic field are detected by magnetization measurements on a system with the number of rings of the order of $10^{11}$ \cite{fominprl}. 
For electrons on the helium surface recent microwave absorption experiments \cite{yunusowa} are performed on  $\sim 10^7$ electrons. 
For our purpose, to eliminate the interaction between the electrons localized above separate tube electrodes, the distances between the tubes need to be large enough. In the present study, the rectangular shape of the computational box does not affect
the circular potential above the tube for the distance of $1.2\mu$m between the axis of the tube and the side of the computational box. The Neumann boundary condition at the sides is justified by the screening of the electron charge by the image charges. Therefore, the distance between the tubes in the array of $2.5\mu$m should be enough to switch off the interaction between the electrons of separate tubes. 
The effects described in this paper deal with a single electron per tube. The electron density in the experiments is tuned by the electric field in the capacitor \cite{kawakami21,yunusowa}. The spacing between the tubes of $2.5\mu$m corresponds to $1.6\times 10^7$cm$^{-2}$, which is nearly equal to the surface electron density in the experimental conditions of e.g. Ref. \cite{yunusowa}.

An experiment with the array may be difficult due to the dependence of the energy positions of the main line of the spectrum on the exact width of the helium layer (Fig. 7), which would require a fine positioning of the container. 
However, the signal from a single tube could be detected with the recent development of the spectroscopy technique based on the capacitive coupling of surface electrons   to the capacitor plates \cite{kawakami19} defining the external electric field
that is indicated for quantum information processing on Rydberg states as qubits.
The method is based on the measurement of pA currents induced by an electron oscillating between the lowest and excited Rydberg states in a resonant microwave field and thus changing its distance from both plates \cite{kawakami19}.

The present modeling neglects the coupling of the electrons to the ripplon surface waves. The effect of the coupling to the ripplon field on the absorption spectra has been considered in Ref.\cite{Dyprb} for a closely related problem of 0D states induced by a rod electrode submerged 500 nm below the surface. The Franck-Condon polaronic shift of the transition frequency between the first and second energy level was estimated as a product of 0.022 GHz and a factor $f_p$ that depends on the ratio between the effective Bohr radius (7.6 nm) and the in-plane confinement length $a_{||}$. For the parameters of Fig. 4, $a_{||}$ is of the order of 100 nm, for which, based on the results of Ref. \cite{Dyprb}, $f_p<0.01$. The ripplonic shift of the energy spectra should be then smaller than 0.22 MHz and thus negligible. The coupling to ripplons shifts the transition lines stronger at higher temperature, but for $T \leq 20$ meV they remain negligible \cite{lambs,lambprl}.

\section{Summary and Conclusions}
We studied ring-like confinement of electrons by 
a tubular electrode placed beneath the surface of liquid helium.
  For the tube that is submerged shallow in helium, the vertical magnetic field induces Aharonov-Bohm angular momentum transitions.
By tuning the helium surface level one can
 arrange for angular momentum transitions occurring at a period of the magnetic field that is significantly different 
for the first and second Rydberg states of the vertical quantization. Then, the absorption spectrum exhibits 
distinct discontinuities at the ground-state angular momentum transitions.
At the temperatures of the order of 10 mK, the discontinuities will no longer be present
since a number of states of various $m$ will be occupied. However, the lines can be resolved
in the energy, and the observation of the intensity of the lines as functions of the external magnetic offers a way for the experimental detection of  Aharonov-Bohm effect for bound electron states.


\section*{Appendix: implementation of the finite element method}
\setcounter{equation}{0} 
\renewcommand{\theequation}{A.\arabic{equation}}
To determine the potential for the electron due to the image charges in
helium and metal electrodes, we solve the Poisson equation using the finite element method.
 We use cubic elements of 100nm side length with spacing between the nearest nodes of 50nm,
 with 27 nodes in each element. Inside the box, we have 8 nodes per element of the boundaries of the system. We work with 60025 elements and 499851 nodes. 

For an element covering $x\in[x_s,x_e]$ a local coordinate $\xi_1\in[-1,1]$ is used with 
\begin{equation} x(\xi_1)=\frac{1-\xi_1}{2}x_s+\frac{1+\xi_1}{2}x_e, \end{equation} and similar mapping for y and z directions with $\xi_2$ and $\xi_3$ local coordinates. Within the element [see Fig. \ref{komorka}] 27 nodes are defined for the
local coordinates $\xi_1,\xi_2$ and $\xi_4$ equal $\pm 1$ or 0.  
We define three parabolic shape functions \begin{eqnarray} \psi_{-1}(\xi)&=&\xi(\xi-1)/2\\\psi_0&=&(1-\xi)(1+\xi) \\\psi_1&=&\xi(\xi+1)/2\end{eqnarray} which are Lagrange node functions with $\psi_n(m)=\delta_{n,m}$ for $n,m=-1,0,1$.
The 3D Lagrange functions for a node with local coordinates $\eta_1,\eta_2,\eta_3$ is defined
as $\Psi_{\eta_1,\eta_2,\eta_3}=\psi_{\eta_1}(\xi_1)\psi_{\eta_2}(\xi_2)\psi_{\eta_3}(\xi_3)$. 
Within the cubic element the potential is spanned in the basis of 27 of these functions
\begin{equation} V(x(\xi_1),y(\xi_2),z(\xi_3))=\sum_{l=1}^{27} c_l \Psi_{\eta_{l1},\eta_{l2},\eta_{l3}}(\xi_1,\xi_2,\xi_3),\end{equation}
where $c_l=V({\bf r}_l)$.
The matrix elements defining the system of equations for potential at the nodes given by Eq.(4) are divided into elements and integrated over the local coordinates.
The three-point differential quotient allows for an exact calculation of the derivatives for
parabolic functions and the Gaussian quadrature provides exact values for integrals of the polynomials.
 The large linear system of equations (about 0.5 mln) is sparse and solved with the PARDISO library.
\setcounter{figure}{0} \renewcommand{\thefigure}{A.\arabic{figure}}
\begin{figure}[htbp]
\centering
\includegraphics[height=0.28\textwidth]{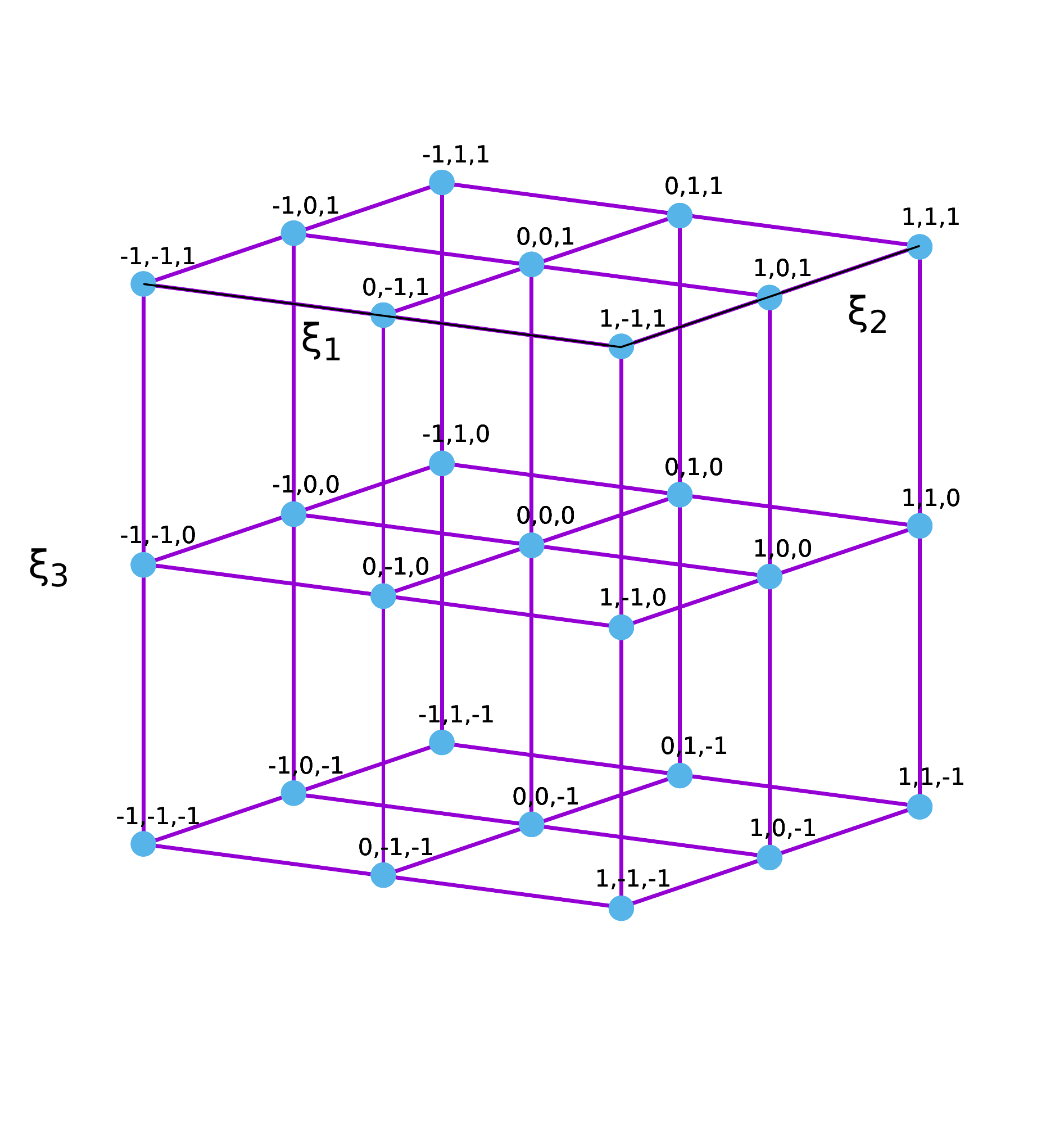} 
\caption{Nodes within a single cubic element in the local system of coordinates $\xi_i\in[-1,1]$
for $i=1,2,3$. The labels close to the nodes indicate the coordinates $(\xi_1,\xi_2,\xi_3)$ of the node.}
\label{komorka}
\end{figure}

The linear system of equations given by Eq. (4) is applied for all nodes inside the computational box.
For the nodes at the sides of the computational box, equations given by boundary conditions are introduced.
Due to the properties of the 
Lagrange functions, the Dirichlet boundary conditions for nodes inside the electrodes can be set  as $c_l=V_{1}$ or $V_2$ 
for nodes $l$ on the bottom or top electrodes, respectively. The Neumann boundary condition for vanishing electric
field component normal to the side faces of the computational box is induced by setting equal values of the coefficients 
$c_{l}=c_{m}$ where $l$ and $m$ are  neighbor nodes adjacent to the boundary in the normal direction. 
With the applied boundary conditions, the surface integral in Eq. (4) is zero. 
At the top and bottom sides of the computational box, i.e., in metal, the potential is constant 
and its gradient vanishes. At the other side of the box, the normal component of the potential is
zero according to the Neumann boundary condition applied therein.

\end{document}